\documentclass[reprint,amsmath,amssymb,aps]{revtex4-1}
\usepackage{graphicx}
\usepackage{dcolumn}
\usepackage{bm}
\usepackage[mathlines]{lineno}

\begin{document}

\preprint{APS/123-QED}

\title{Anisotropic structural and optical properties of $a$-plane $(11\bar{2}0)$ AlInN nearly-lattice-matched to GaN}

\author{Masihhur R. Laskar$^{1}$}
\email[]{laskar@tifr.res.in}
\author{Tapas Ganguli$^{2}$}
\author{A. A. Rahman$^{1}$}
\author{Ashish Arora$^{1}$}
\author{Nirupam Hatui$^{1}$}
\author{M. R. Gokhale$^{1}$}
\author{Sandip Ghosh$^{1}$}
\author{Arnab Bhattacharya$^{1}$}

\affiliation{$^{1}$DCMP\&MS, Tata Institute of Fundamental Research, Mumbai 400005, India.\\ $^{2}$Raja Ramanna Center for Advanced Technology, Indore 425013, India.}


\begin{abstract}

We report epitaxial growth of $a$-plane $(11\bar{2}0)$ AlInN layers nearly-lattice-matched to GaN. Unlike for $c$-plane oriented epilayers, $a$-plane Al$_{1-x}$In$_{x}$N cannot be simultaneously lattice-matched to GaN in both in-plane directions. We study the influence of temperature on indium incorporation and obtain nearly-lattice-matched Al$_{0.81}$In$_{0.19}$N at a growth temperature of $760^{o}C$. We outline a procedure to check in-plane lattice mismatch using high resolution x-ray diffraction, and evaluate the strain and critical thickness. Polarization-resolved optical transmission measurements of the Al$_{0.81}$In$_{0.19}$N epilayer reveal a difference in bandgap of $\sim$$140$ meV between (electric field) $\textbf{E}$$\parallel$$\textbf{c}$ $[0001]$-axis and $\textbf{E}\bot \textbf{c}$ conditions with room-temperature photoluminescence peaked at $3.38$ eV strongly polarized with $\textbf{E}\parallel \textbf{c}$, in good agreement with strain-dependent band-structure calculations.

\end{abstract}

\pacs{Valid PACS appear here}

\maketitle

In the [0001] $c$-plane orientation, AlInN (In$\sim$$18$\%) can be perfectly lattice matched to GaN. This, along with a large bandgap and high refractive index contrast with GaN makes it ideal for Bragg mirrors, microcavity structures \cite{butte,doraz,carlin}, strain-free transistors, photodetectors, and emitters \cite{Medjdoub,senda}. However, to minimize the deleterious effect of spontaneous and piezoelectric polarization fields, growing nitrides along non-polar orientations has advantages, and enables polarization-sensitive optical devices \cite{paskova}. Epitaxial growth of AlInN is extremely challenging since the optimum growth conditions for AlN and InN are very different, the alloy tends to phase separate and shows composition inhomogeneities \cite{Matsuoka,minj,hums}. Optimized growth of AlInN along the $[0001]$ $c$-axis orientation is reported \cite{doraz,carlin,Medjdoub,senda,hums,fuji}. However, the growth mechanisms for non-polar oriented epilayers are different and much less studied. Unlike the $c$-plane case, $a$-plane AlInN can not be grown perfectly lattice-match to GaN. In this letter, we report the synthesis of nearly-lattice-matched $a$-plane $(11\bar{2}0)$ Al$_{0.81}$In$_{0.19}$N via metalorganic vapor phase epitaxy (MOVPE). The anisotropic structural properties are evaluated from high-resolution x-ray diffraction (HRXRD), and strain and critical thickness estimated. The measured optical properties show a strong polarization anisotropy and are compared with results from band-structure calculations.


All epilayers were grown via MOVPE on $r$-plane sapphire in a close-coupled showerhead reactor using standard precursors. Structural characterization was carried out using a Philips X'PERT$^{TM}$ HRXRD system. Optical transmission measurements were performed on backside-polished samples using a Glan-Taylor polarizer on a Cary $5000$ spectrophotometer. Photoluminescence (PL) measurements were performed using 266 nm laser excitation. The surface morphology was studied by optical and atomic force microscopy.


Fig.1 shows the position of relaxed AlN, GaN, InN and sapphire in lattice-parameter space. The $c/a$ ratio of GaN (1.626) is a little higher than that of  AlN (1.600) and InN (1.603) \cite{angerer,maleyre}, hence for $(11\bar{2}0)$ oriented epilayers it is impossible to get perfect lattice matching of AlInN on GaN along both in-plane directions. The inset of Fig.1 shows the relative orientation of the AlInN and GaN unit cells. From Vegard's Law it can be shown that the Al$_{1-x}$In$_{x}$N/GaN system will be lattice-matched along $\textbf{m}$-axis ($\parallel\textbf{y}$) for $x$=$0.18$ (i.e. $OB$=$O_1B_1$, but $BC$$<$$B_1C_1$), whereas lattice-matching along $\textbf{c}$-axis($\parallel\textbf{z}$) requires $x$=$0.28$ (i.e. $BC$=$B_1C_1$, but $OB$$>$$O_1B_1$).

\begin{figure}[b]
\centerline{\includegraphics*[width=6cm]{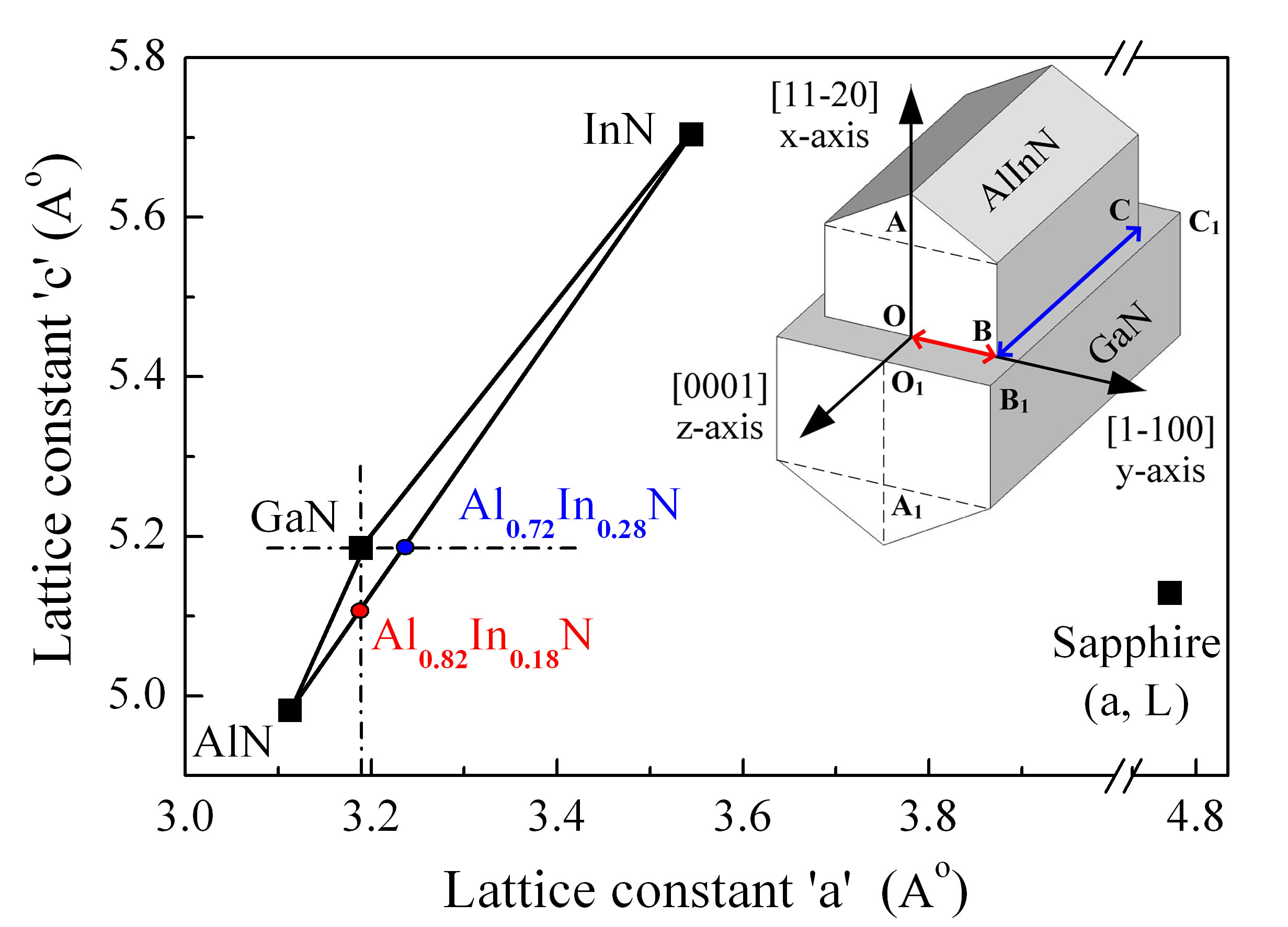}}
\caption{\label{fig1} Lattice parameter space showing the positions of relaxed AlN, GaN, InN and sapphire \cite{ftnote}. Inset: Relative orientation of $a$-plane AlInN unit cell with respect to GaN. For perfect lattice matching, $OB$=$O_1B_1$ and $BC$=$B_1C_1$ simultaneously which is  not possible at any single composition.}
\end{figure}

The indium incorporation into AlInN ($x_{solid}$) depends strongly on growth temperature, and was studied by growing a set of AlInN epilayers on AlN buffer layers over the temperature range $600$$-$$1040^{o}C$ at fixed $x_{gas}$(TMIn/(TMAl+TMIn))=$0.40$. The $x_{solid}$ was estimated from the $(11\bar{2}0)$, $(10\bar{1}0)$, $(10\bar{1}1)$, $(10\bar{1}2)$ and $(11\bar{2}2)$ reflections following the procedure in Ref\cite{mrlLattice} (with $\pm$1-2\% error). As shown in Fig.2, the $x_{solid}$ is close to the $x_{gas}$-value at $600^{o}C$. With increasing temperature the desorption of In-atoms from the surface increases, hence reducing solid phase In-incorporation. At high temperature, $\geq$$900^{o}C$, typically needed for high quality Al-alloys, $x_{solid}$ is less than $3$\%, showing the difficulty of growing AlInN. To obtain $x_{solid}$$\approx$18\% (lattice matched to GaN along $\textbf{y}$-axis), the temperature required is $\sim$$760^{o}C$.

\begin{figure}[t]
\centerline{\includegraphics*[width=6cm]{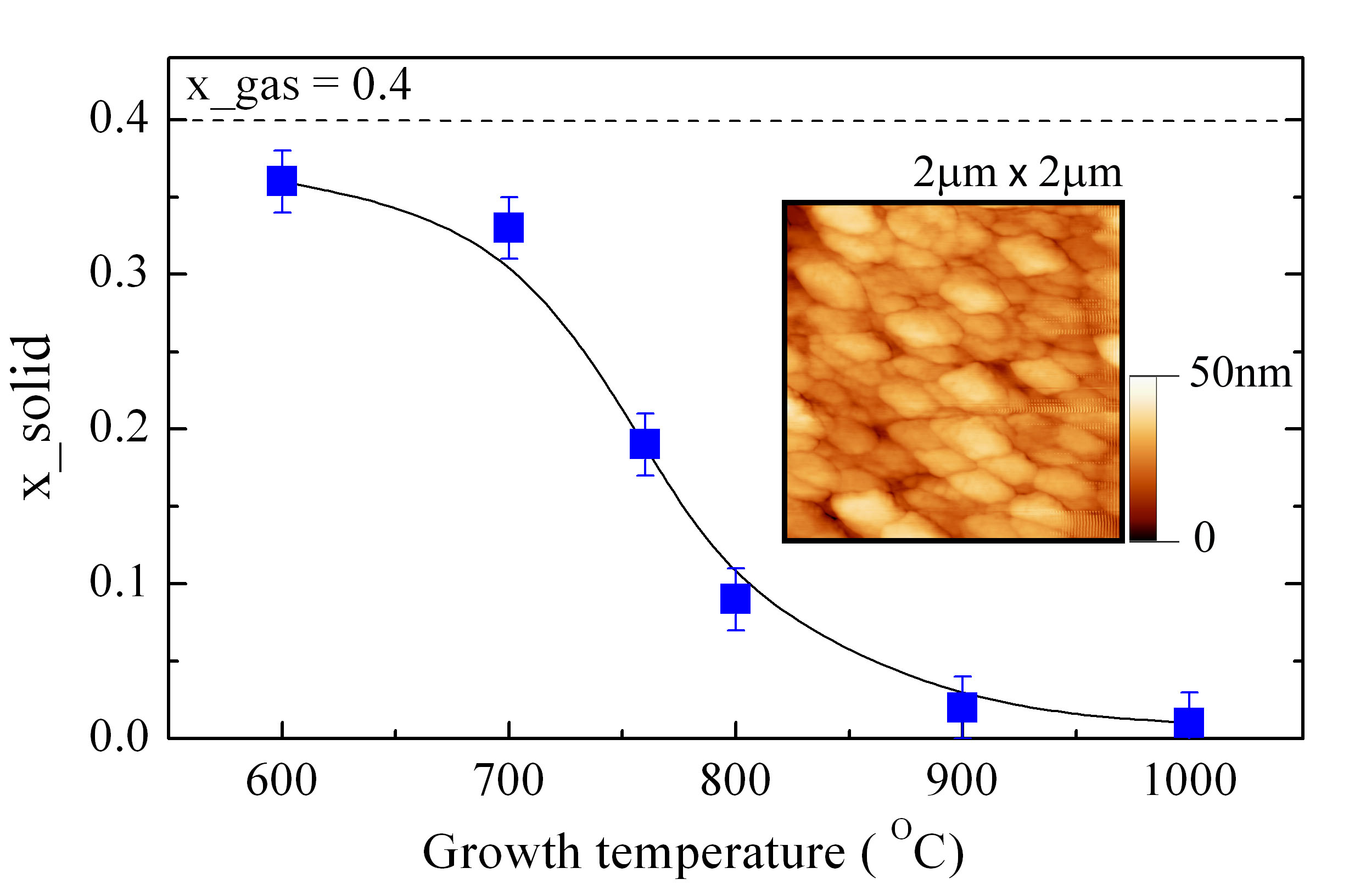}}
\caption{\label{fig1} The variation of $x_{solid}$ as a function of growth temperature where $x_{gas}$=$40$\% fixed. Solid line is a guide to the eye. Inset: the AFM image of the Al$_{0.81}$In$_{0.19}$N epilayer grown on GaN buffer with an RMS roughness of $\sim 6 nm$.}
\end{figure}

\begin{figure}[b]
\centerline{\includegraphics*[width=7.5cm]{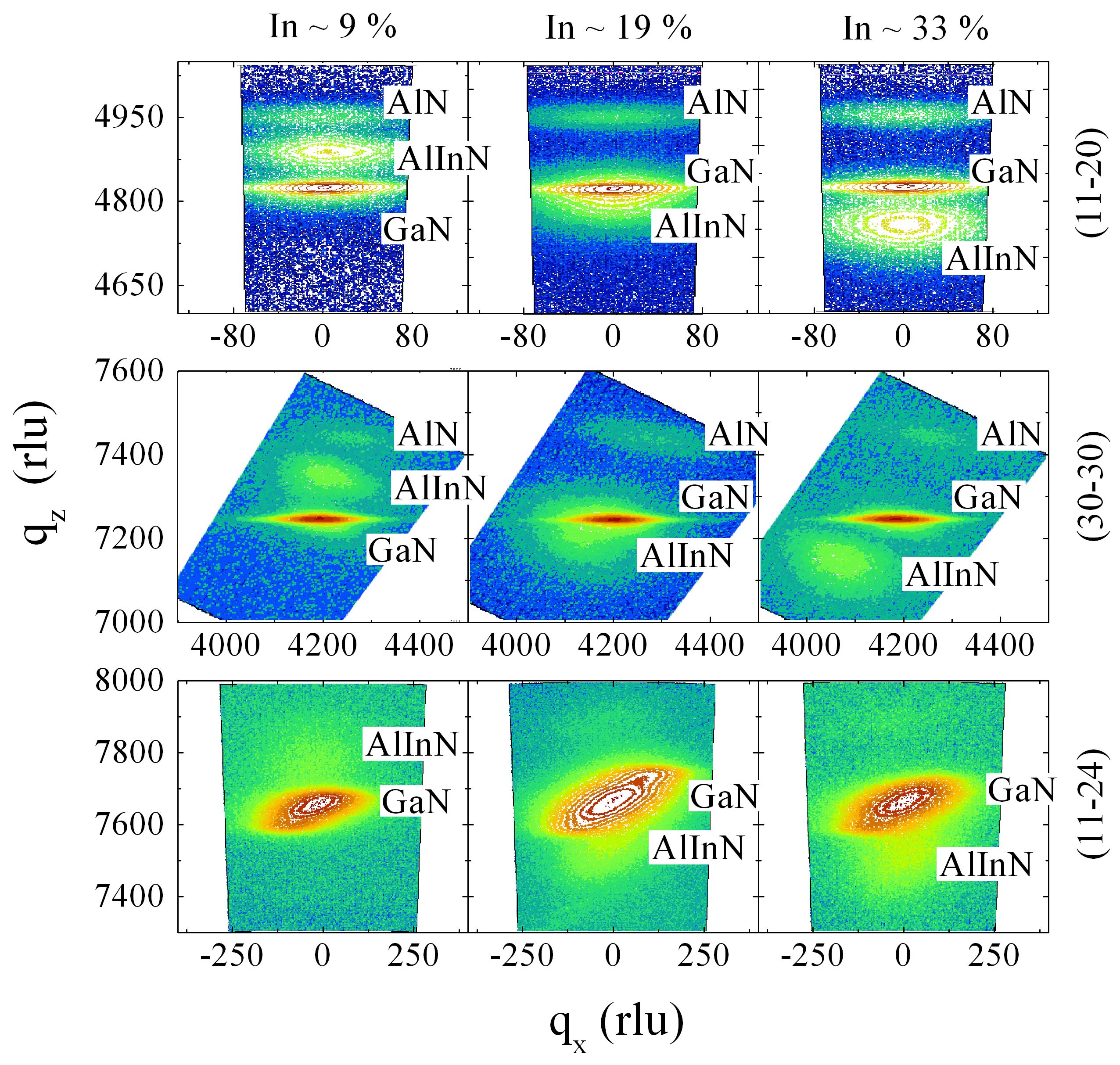}}
\caption{\label{fig1} The matrix of reciprocal space maps of $(11\bar{2}0)$, $(30\bar{3}0)$, and $(11\bar{2}4)$ reflections of AlInN epilayer grown on GaN at $770^{o}C$, $760^{o}C$, and $800^{o}C$ having $x_{solid}$ of $\sim$$9$\%, $\sim$$19$\% and $\sim$$33$\% respectively. Here  $1$ $rlu$(reciprocal lattice unit)=$\lambda/2d$, $\lambda$=X-ray wavelength and $d$=(hkil) interplaner distance.}
\end{figure}

Fig.3 shows a set of reciprocal space maps (RSM) of AlInN epilayers deposited on GaN buffer layer at $700^{o}C$, $760^{o}C$ and $800^{o}C$ having $x_{solid}$ of $\sim$$9$\%, $\sim$$19$\% and $\sim$$33$\% respectively. Our diffractometer cannot directly probe the $(0001)$ and $(1\bar{1}00)$ reflections; we hence rely on a combination of symmetric and asymmetric reflections to confirm lattice matching (LM) along different in-plane directions. For an $a$-plane oriented layer the $(11\bar{2}0)$ plane is symmetric, hence its lattice point (LP) is on the vertical axis in reciprocal space. To verify LM of AlInN/GaN along the $\textbf{m}$-axis (Inset:Fig.1), we use the $(30\bar{3}0)$-plane, whose LP lies in a plane both $\bot$ to the $\textbf{c}$-axis and $\parallel$ to the $\textbf{m}$-axis in reciprocal space. The RSMs shows that the diffraction peak of Al$_{0.81}$In$_{0.19}$N epilayer coincides with GaN for the symmetric $(11\bar{2}0)$ and asymmetric $(30\bar{3}0)$ reflections, confirming LM along $\textbf{m}$-axis. Similarly to verify LM along the $\textbf{\textbf{c}}$-axis, we use the $(11\bar{2}4)$ reflection, since its reciprocal lattice point lies in a plane both $\bot$ to the $\textbf{m}$-axis and $\parallel$ to the $\textbf{c}$-axis. The separate peaks of AlInN and GaN indicate that the layer is not LM along the $\textbf{c}$-axis.

\begin{figure}[b]
\centerline{\includegraphics*[width=7cm]{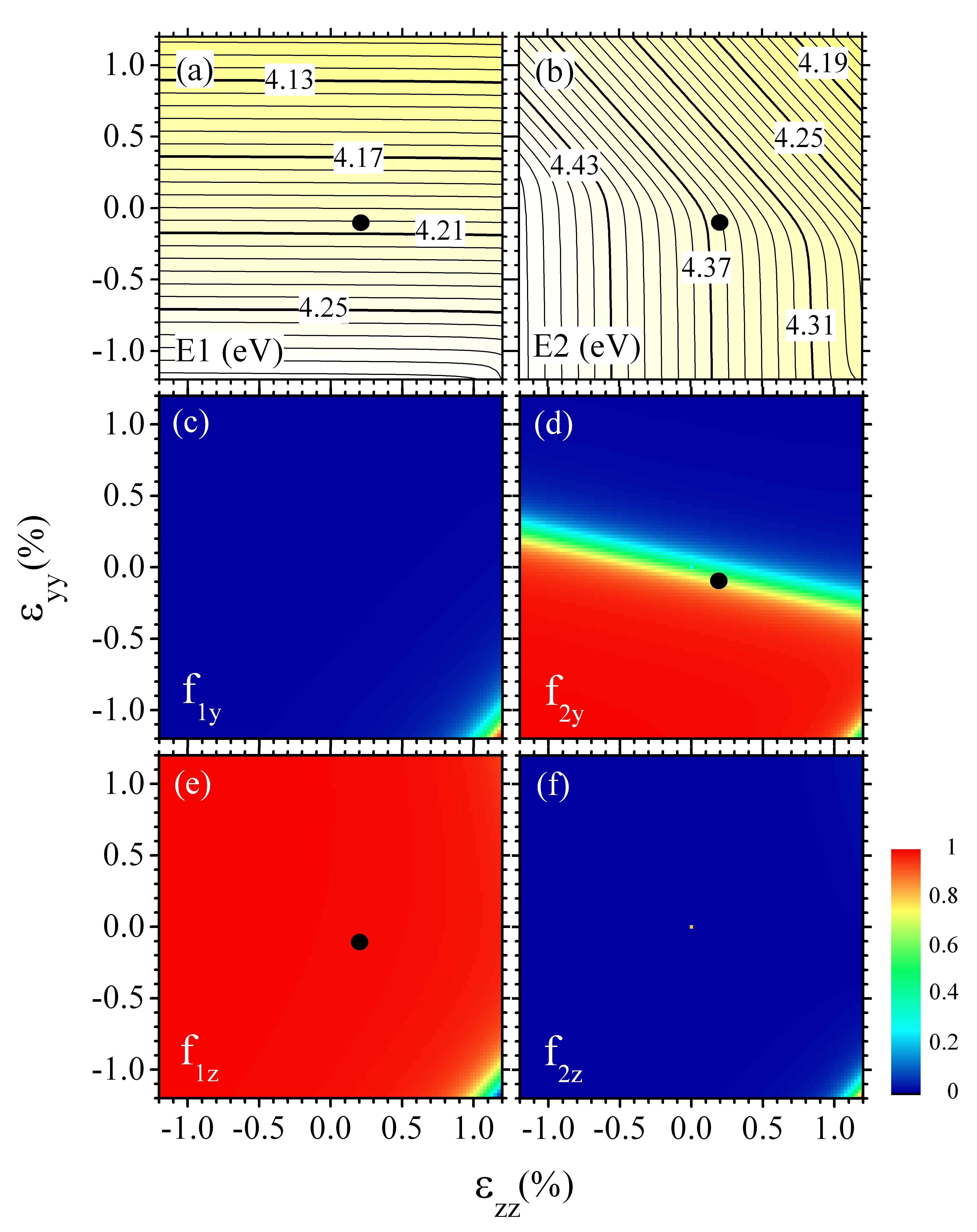}}
\caption{\label{fig1} The calculated transition energies (a) $E1$, (b) $E2$ are shown in the strain parameter space of $\epsilon_{yy} $ and $\epsilon_{zz}$. (c) and (e) shows the relative oscillator strength of $E1$-transition energy for \textbf{y} and \textbf{z}-polarization respectively. (d) and(f) are same for $E2$-transition energy.}
\end{figure}

The advantage of growing Al$_{0.81}$In$_{0.19}$N on GaN (LM along $\textbf{m}$-axis) rather than Al$_{0.72}$In$_{0.28}$N (LM along $\textbf{c}$-axis) is that it provides higher refractive index contrast making it more useful for Bragg mirrors. Although Al$_{0.81}$In$_{0.19}$N on GaN has nearly zero lattice misfit along $\textbf{m}$-axis, the strain due to lattice-misfit $f_{z}$=$-1.42$\% along $\textbf{c}$-axis can cause cracking beyond a certain critical thickness ($t_c$). A theoretical estimate \cite{matthews,murray} of $t_c$ for Al$_{0.81}$In$_{0.19}$N on GaN is $\approx$$250$nm which is relatively smaller than the $\geq$$700$nm thickness at which we typically observe cracks.

To study the optical properties we use Al$_{0.81}$In$_{0.19}$N epilayers grown on transparent AlN buffer layers and measure the bandgap from polarization-resolved optical transmission and PL measurements. Wurtzite group III-nitrides have intrinsic polarization anisotropy in their optical properties \cite{sandip}. In epitaxial alloys, the composition and in-plane strain causes the mixing of the three upper-most valance bands, which not only changes the transition energies but also modifies the polarization selection rules for each transition. For $x_{solid}$$=$$0.19$, we have calculated the three lowest interband transition energies ($E1$, $E2$, $E3$) and their oscillator strengths (OS) for \textbf{x}, \textbf{y}, \textbf{z}-polarization by solving the \emph{Bir-Pikus} Hamiltonian \cite{ftnote2}. Fig.4 shows the calculated $E1$, $E2$ values and their \textbf{y}, \textbf{z} component of OS as a function of $\epsilon_{yy}$ and $\epsilon_{zz}$. Using HRXRD we estimate the in-plane anisotropic strain in our layer as $\epsilon_{yy}$=$-0.1$\% and $\epsilon_{zz}$=$+0.2$\% (black dots in Fig.4[d-e]), for which $E1$ transition is strongly \textbf{z}-polarized and $E2$ transition is predominantly \textbf{y}-polarized.

\begin{figure}[t]
\centerline{\includegraphics*[width=6cm]{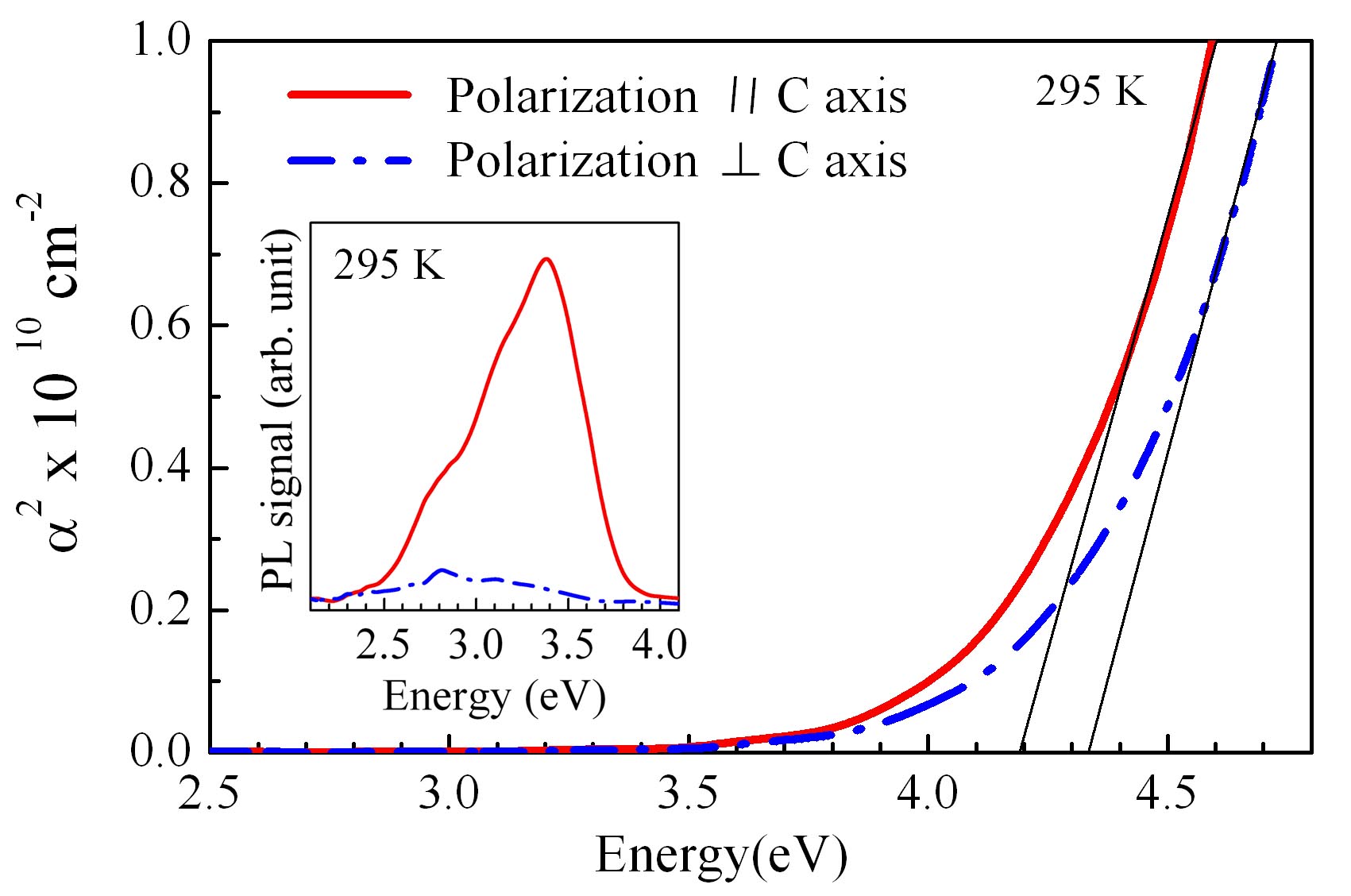}}
\caption{\label{fig1} The absorption spectra of Al$_{0.19}$In$_{81}$N epilayer showing the bandgap $ \sim 4.20$ $eV$ and $ \sim 4.34$ $eV$ for $\mathbf{E}\parallel \mathbf{c}$ and $\mathbf{E}\bot \mathbf{c}$ polarization respectively. Inset: Room temperature PL from the same epilayer showing that the emission peak at $3.38$ $eV$ strongly polarized with $\mathbf{E}\parallel \mathbf{c}$.}
\end{figure}

Fig.5 shows the absorption spectra for two different polarizations (electric field) $\mathbf{E}\parallel \mathbf{c}$ ($\parallel \textbf{z}$) and $\mathbf{E}\bot \mathbf{c}$ ($\parallel \textbf{y}$) for which the extrapolated bandgaps are $\sim$$4.20$ eV and $\sim$$4.34$ eV respectively (which we had assigned as $E1$ and $E2$ in our calculation). The measured difference in bandgap $\triangle E$=$E2$$-$$E1$$\approx$$140 $ meV agrees well with the theoretical estimate $\sim$$155$ meV (from Fig.4a\&b). The Fig.5 inset shows the room temperature PL spectrum which peaks at $\sim$$3.38$ eV. Such a large Stokes-shift of $\sim$$0.85$eV in PL arises because of indium localization and is similar to values observed for $c$-plane epilayers \cite{butte,zhou}. The PL spectrum is strongly polarized at $\mathbf{E}\parallel \mathbf{c}$ as expected because the lowest energy transition $E_1$ has a strong \textbf{z}-polarized OS and very weak \textbf{y}-polarized OS. The UV bandgap of Al$_{0.81}$In$_{0.19}$N coupled with its inherently anisotropic optical properties makes it promising for polarization-sensitive detectors and emitters in the UV.

In conclusion, we show it is impossible to achieve perfectly-lattice-matched $(11\bar{2}0)$ AlInN on GaN and suggest a method to verify in-plane lattice mismatch. A study of In-incorporation as a function of temperature allows us to obtain nearly-lattice-matched Al$_{0.81}$In$_{0.19}$N at $760^{o}C$. Polarization-resolved optical measurements show a strong anisotropy. Our results would be useful for polarization-sensitive detectors and polarized emitters in the UV, and non-polar AlInN-based Bragg mirrors.

%


\end{document}